\documentclass[preprint,12pt]{elsarticle}

\usepackage{amssymb}
 \usepackage{amsthm}

\journal{Nuclear Physics B}

\begin{document}

\begin{frontmatter}



\title{Defect branes}


\author[ea]{Eric A. Bergshoeff}

\ead{E.A.Bergshoeff@rug.nl}

\author[to]{Tomas Ort\'\i n}

\ead{Tomas.Ortin@csic.es}

\author[fr]{Fabio Riccioni}

\ead{fabio.riccioni@roma1.infn.it}

\address[ea]{Centre for Theoretical Physics, University of Groningen, \\ Nijenborgh 4, 9747 AG Groningen, The
Netherlands}

\address[to]{Instituto de F\'\i sica Te\'orica UAM/CSIC \\
C/ Nicol\'as Cabrera, 13-15, C.U. Cantoblanco, E-28049-Madrid, Spain}

\address[fr]{INFN Sezione di Roma, \\  Dipartimento di Fisica, Universit\`a di Roma ``La Sapienza'' \\ Piazzale Aldo Moro 2, 00185 Roma, Italy}

\begin{abstract}
  We discuss some general properties of ``defect branes'', i.e.~branes of
  co-dimension two, in (toroidally compactified) IIA/IIB string theory. In
  particular, we give a full classification of the supersymmetric defect
  branes in dimensions $3\le D\le 10$ as well as their higher-dimensional
  string and M-theory origin as branes and a set of ``generalized''
  Kaluza-Klein monopoles. We point out a relation between the generalized
  Kaluza-Klein monopole solutions and a particular type of mixed-symmetry
  tensors.  These mixed-symmetry tensors can be defined at the linearized
  level as duals of the supergravity potentials that describe propagating
  degrees of freedom. It is noted that the number of supersymmetric defect
  branes is always twice the number of corresponding central charges in the
  supersymmetry algebra.

\end{abstract}


\end{frontmatter}


\section{Introduction}
\label{introduction}

Branes are a fundamental ingredient of string theory. Prime examples of their
many applications are the calculation of the entropy of certain black holes
\cite{Strominger:1996sh} and the AdS/CFT correspondence
\cite{Maldacena:1997re}. The properties of branes crucially depend on two
quantities: the scaling of the brane tension with the string coupling constant
$g_s$ in the string frame and the number $T$ of transverse directions. The
first quantity can be characterized by a number $\alpha$ such that
\begin{equation}
{\rm Tension} \ \ \sim\ \ (g_s)^\alpha\,.
\end{equation}
It turns out that $\alpha$ is a non-positive number.\footnote{We do not
  consider instantons here. They will be shortly discussed in the conclusion
  section.}  Branes with $\alpha=0\,, -1\,, -2\,, \dots$ are called
Fundamental\,, Dirichlet\,, Solitonic\,, etc. The second quantity $T$
naturally splits the branes into two classes: the {\sl standard} branes with
$T\ge 3$ and the {\sl non-standard} ones with $T=2,1,0$.  Only the standard
branes are asymptotically flat. The non-standard branes require special
attention. For instance, the non-standard branes with $T=0$ are space-filling
branes which can only be defined consistently in combination with as
orientifold.  The ones with $T=1$ are domain walls. The potentials coupling to
these domain walls are dual to constants such as mass parameters
or gauge coupling constants. By T-duality, these domain walls need
orientifolds as well \cite{Polchinski:1995df}.

In this paper we wish to focus on non-standard branes with $T=2$. We call such
branes ``defect branes'' since branes with co-dimension 2, like the D7-brane
or 4D cosmic strings, are not asymptotically flat and can have non-trivial
deficit angles at spatial infinity. A prime example of a Dirichlet defect
brane is the ten-dimensional D7-brane \cite{Polchinski:1995mt} whose solution
has been discussed in
\cite{Greene:1989ya,Gibbons:1995vg,Bergshoeff:2006jj}. It is well-known that
the {\sl single} D7-brane solution has no finite energy
\cite{Greene:1989ya,Gibbons:1995vg}. To obtain such a finite-energy solution
one should construct a {\sl multiple} brane solution which includes
orientifolds. In this paper we will only consider {\sl single} defect branes
and assume that finite energy solutions can be obtained by applying the same
techniques as for the D7-brane.

Defect branes couple to $(D-2)$-form potentials.  These potentials are dual to
the ${\rm dim}\, G - {\rm dim}\, H$ scalars that para\-me\-trize the
non-linear coset $G/H$ of the corresponding maximal supergravity
theory.\,\footnote{We note that it is non-trivial to use these potentials to
  describe {\sl multiple} defect branes. For instance, it is not clear how to
  express the branch-cuts of the holomorphic axion-dilaton solution in terms of
  properties of the corresponding dual potentials. We thank Jelle Hartong for
  a discussion on this point.}  It turns out that the number $n_{\rm P}$ of
$(D-2)$-form potentials is not equal to the number $n_{{\rm S}}$ of coset
scalars, i.e.~$n_{\rm P} \ne n_{\rm S}$, see Table 1.  The reason of this is
that the $(D-2)$-form potentials transform in the adjoint representation of
the duality group $G$.  Their $(D-1)$-form field strengths are essentially the
Hodge duals of the Noether current 1-forms associated to the global invariance
under $G$ \cite{Bergshoeff:2009ph}, which transform in the adjoint
representation of $G$. The ${\rm dim}\,G$ Noether currents are constrained by
${\rm dim}\, H$ relations \cite{modlt} and, therefore, the $(D-2)$-form
potentials describe as many physical degrees of freedom as the coset scalars.
These constraints, however, do not lead to algebraic relations among the
potentials themselves and therefore do not play a role in the present
discussion.

To determine whether we are dealing with a {\sl supersymmetric} defect brane
we will use a criterion that is based on the construction of a gauge-invariant
Wess-Zumino (WZ) term that describes the coupling of the defect brane to a
given $(D-2)$-form potential \cite{Bergshoeff:2006gs,Bergshoeff:2010xc}. This
WZ term should contain worldvolume fields that precisely fit into a
half-supersymmetric vector or tensor multiplet.  This supersymmetric brane
criterion leads to a full classification of supersymmetric defect branes in
dimensions $3 \le D \le 10$. It turns out that the number $n_{{\rm D}}$ of
supersymmetric defect branes in any dimension is less than the number $n_{\rm
  P}$ of $(D-2)$-form potentials, i.e.~$n_{\rm D} < n_{\rm P}$. This means
that not all potentials correspond to supersymmetric branes, see Table 1.
This is different from the standard branes where the number of potentials
always equals the number of supersymmetric branes.  The number of {\it all}
non-standard branes have been recently derived in dimension higher than five
in \cite{Bergshoeff:2011qk} using the method of \cite{Bergshoeff:2010xc}, and
in {\it all} dimensions in \cite{Kleinschmidt:2011vu} using an approach based
on ${\rm E}_{11}$ \cite{West:2001as} and the observation that imaginary roots
do not lead to supersymmetric branes \cite{Houart:2011sk}.  As far as the
number $n_{\rm D}$ of defect branes is concerned, we will give yet another
derivation of this number using a different method, see Section 2.  The final
result can be found in Table \ref{table1}.  This Table also shows that in
$D<10$ the number $n_{{\rm D}}$ of supersymmetric defect branes is not equal
to the number $n_{{\rm S}}$ of coset scalars, i.e. $n_{\rm D} \ne n_{\rm S}$.
It is just a coincidence that these two numbers are the same in ten
dimensions.

\begin{table}[h]
\begin{center}
\begin{tabular}{|c|c|c|c|c|c|c|}
\hline\rule[-1mm]{0mm}{6mm}
$D$&$G/H$&$n_{\rm P}$&$n_{\rm D}$&$n_{\rm S}$\\[.1truecm]
\hline \rule[-1mm]{0mm}{6mm} {\rm IIA}&--&--&--&--  \\[.05truecm]
\hline \rule[-1mm]{0mm}{6mm} {\rm IIB}&${\rm SL}(2,\mathbb{R})/{\rm SO}(2)$&3&2& 2  \\[.05truecm]
\hline \rule[-1mm]{0mm}{6mm}9 &${\rm SL}(2,\mathbb{R})/{\rm SO}(2)\times \mathbb{R}^+$&4&2&3    \\[.05truecm]
\hline \rule[-1mm]{0mm}{6mm}8 &${\rm SL}(3,\mathbb{R})/{\rm SO}(3)\times {\rm SL}(2,\mathbb{R})/{\rm SO}(2)$&11&8&7    \\[.05truecm]
\hline \rule[-1mm]{0mm}{6mm}7 &${\rm SL}(5,\mathbb{R})/{\rm SO}(5)$&24&20&14    \\[.05truecm]
\hline \rule[-1mm]{0mm}{6mm}6&${\rm SO}(5,5)/{\rm SO}(5)\times{\rm SO}(5)$&45&40& 25     \\[.05truecm]
\hline \rule[-1mm]{0mm}{6mm}5 &${\rm E}_6/{\rm Sp}(8)$&78&72&42    \\[.05truecm]
\hline \rule[-1mm]{0mm}{6mm}4 &${\rm E}_7/{\rm SU}(8)$&133&126& 70  \\[.05truecm]
\hline \rule[-1mm]{0mm}{6mm}3 &${\rm E}_8/{\rm SO}(16)$&248&240& 128  \\[.05truecm]
\hline
\end{tabular}
\end{center}
 \caption{\sl Comparison between the number $n_{\rm P} = {\rm dim}\, G$ of $(D-2)$-form potentials,
 the number $n_{\rm D}=
 {\rm dim}\, G - {\rm rank}\, G$ of supersymmetric defect branes
and the
 number $n_{\rm S}= {\rm dim}\,
G - {\rm dim}\, H$ of coset scalars
for the  coset spaces $G/H$ of  maximal supergravity in $3 \le D \le 10$
dimensions. The derivation of the expression for $n_{\rm D}$ may be found
in Section~\ref{sec-defectbranes}.}\label{table1}
\end{table}

The lower-dimensional branes with $\alpha = 0,-1,-2,-3$ can all be seen to
arise as dimensional reductions of branes and a set of generalized KK
monopoles in ten dimensions \cite{Bergshoeff:2011mh,Bergshoeff:2011ee}. The
generalized KK monopoles can be schematically represented by the introduction
of mixed-symmetry fields in ten dimensions, provided that one applies a
restricted dimensional reduction rule when counting the branes in the lower
dimension: given a mixed-symmetry field $A_{m,n}$ with $m>n$, indicating a
Young tableaux consisting of a column of length $m$ and a column of length
$n$, one requires that the $n$ indices have to be internal and parallel to $n$
of the $m$ indices \cite{Bergshoeff:2011mh,Bergshoeff:2011ee}.\,\footnote{
  This rule naturally generalizes to the case of fields with more than two
  sets of antisymmetric indices corresponding to a Young tableaux with more
  than 2 columns \cite{Bergshoeff:2011ee}.} Here we generalize this result,
and we determine all the ten-dimensional mixed-symmetry fields that are
required to generate all the defect branes for any value of $\alpha$ using the
restricted reduction rule. We also derive the eleven-dimensional origin of
these fields.

Remarkably, all the solutions corresponding to the generalized KK mono\-poles
that we introduce here were already determined in
\cite{LozanoTellechea:2000mc}, and as we will show the restricted reduction
rule automatically translates into the dictionary used in
\cite{LozanoTellechea:2000mc} to classify these solutions. The mixed-symmetry
fields we introduce can all be seen as generalized duals
\cite{Curtright:1980yk,Hull:2001iu} of the graviton and the other potentials
in the ten- or eleven-dimensional theory. This means that at least at the
linearized level one can impose a duality relation, which can be used to
predict the behaviour of the fields in the various solutions. By explicitly
writing down some of the explicit solutions of \cite{LozanoTellechea:2000mc},
we will show that this predicted behaviour is indeed correct.

The organization of this paper is as follows. In
Section~\ref{sec-defectbranes} we derive the expression for the number $n_{\rm
  D}$ of supersymmetric defect branes given in Table \ref{table1}. In Section
3 we give the string and M-theory origin of these defect branes in terms of
branes and a set of ``generalized'' Kaluza-Klein (KK) monopoles. Furthermore,
we discuss the relation between the generalized KK monopoles and
mixed-symmetry fields of a certain type.  In Section~\ref{sec-mixedsymemtry}
we show how these mixed-symmetry fields classify all defect brane solutions.
As an example we give the string and M-theory monopole solutions that give
rise to all the $D=8$ defect branes. We also show in Section
\ref{sec-explicitsolutions} how the linearized duality relations between these
mixed-symmetry fields and the propagating forms determine the behaviour of the
fields in the various solutions. This is compared with the explicit known
results in all cases.  In Section~\ref{sec-centralcharges} we explain why the
number $n_{\rm D}$ of supersymmetric defect branes is, for each dimension $D$,
equal to twice the number $n_{\rm Z}$ of corresponding central charges in the
supersymmetry algebra. In the final Section we give our conclusions.


\section{Supersymmetric defect branes}
\label{sec-defectbranes}

At first sight one might think that the number $n_{\rm D}$ of supersymmetric
defect branes is equal to the number $n_{\rm P}$ of dual $(D-2)$-form
potentials.  However, this is not the case. A prime example is ten-dimensional
IIB string theory where the 8-forms are in the ${\bf 3}$ of ${\rm
  SL}(2,\mathbb{R})$ and we only have a supersymmetric D7-brane and its
S-dual, i.e.~$n_{\rm D}=2$ \cite{Bergshoeff:2006gs}. The reason why we only
have two supersymmetric seven-branes can be seen as follows: using an ${\rm
  SO}(2,1)$ notation the WZ terms for the three candidate seven-branes can be
written in a duality-covariant way schematically as follows:
\begin{equation}\label{WZ} {\rm WZ}_i \ \ \sim \ A_{8,i} \ + \ \ {\overline
    {{\cal
F}}_{2}}\,\Gamma_i\, A_{6}\ +\ \dots\,,\hskip 1truecm i=+,-,3\,,
\end{equation}
where we have used lightcone notation to label the ${\rm SO}(2,1)$ gamma
matrices $\Gamma_i$.  Here ${\cal F}_2$ is a 2-component spinor\,\footnote{We
  do not write explicitly the spinor indices here.} of ${\rm SO}(2,1)$ whose
components are the worldvolume curvatures of the Born-Infeld vector and its
S-dual. Similarly, the target-space potentials $A_6$ are a spinor (doublet) of
${\rm SO}(2,1)$ whose components are the NS-NS and RR 6-form potentials.  In
general the above expression (\ref{WZ}) for the WZ term contains {\sl two}
worldvolume vectors which do not fit into a single vector multiplet. Therefore
we need that, for a given value of the index $i$ the gamma matrices $\Gamma_i$
act as a projection operator that projects out one of the two worldvolume
vectors in the expression (\ref{WZ}). It turns out that this is the case for
$i=+$ and $i=-$ but not for $i=3$. This explains why there is no
supersymmetric solitonic ($\alpha=-2$) seven-brane in ten dimensions.

We now consider the counting of supersymmetric defect branes in $D<10$
dimensions. We first decompose the adjoint of the U-duality group $G$ under
the direct product of the T-duality group $T={\rm SO}(10-D,10-D)$ and a
scaling symmetry $\mathbb{R}^+$ of the {\sl $D$-dimensional} string coupling
constant.  We find that for each dimension $D\ge 5$ this adjoint
representation decomposes into a Dirichlet, i.e.~$\alpha=-1$, spinor of
T-duality with real components, a Solitonic adjoint plus singlet of T-duality
and a charge-conjugate spinor of T-duality with $\alpha=-3$:
\begin{equation}\label{decomposition}
{\rm Adj}_{|U} = {\rm spinor}_{\alpha=-1}\ +\ \big({\rm Adj}_{|T} +
{\rm singlet}\big)_{\alpha=-2}\ +\  {\rm (conj.\
spinor)}_{\alpha=-3}\,.
\end{equation}
In four and three dimensions this decomposition is modified, and one gets
\begin{eqnarray}
& &  {\rm Adj}_{|{\rm E}_7} = {\rm singlet}_{\alpha=0} +  {\rm spinor}_{\alpha=-1}\ +\ \big({\rm Adj}_{|T} +
{\rm singlet}\big)_{\alpha=-2}\ \nonumber \\
& &  +\  {\rm (conj.\
spinor)}_{\alpha=-3} + {\rm singlet}_{\alpha=-4}  \label{decompositionD=4}
\end{eqnarray}
in four dimensions and
\begin{eqnarray}
& &  {\rm Adj}_{|{\rm E}_8} = {\rm vector}_{\alpha=0} +  {\rm spinor}_{\alpha=-1}\ +\ \big({\rm Adj}_{|T} +
{\rm singlet}\big)_{\alpha=-2}\ \nonumber \\
& &  +\  {\rm (conj.\
spinor)}_{\alpha=-3} + {\rm vector}_{\alpha=-4}  \label{decompositionD=3}
\end{eqnarray}
in three dimensions.

The non-standard branes with $\alpha=-1$, $\alpha =-2$ and $\alpha=-3$ have
been classified in \cite{Bergshoeff:2010xc}, \cite{Bergshoeff:2011zk} and
\cite{Bergshoeff:2011ee} respectively. By looking at equation
(\ref{decomposition}), this implies the classification of all defect branes in
any dimension above four. Moreover, the $\alpha=-4$ branes in four and three
dimensions can easily be obtained by the S-duality\,\footnote{We are referring
  here to the {\it $D$-dimensional} S-duality in which the $D$-dimensional
  string coupling constant (the exponential of the $D$-dimensional dilaton)
  is inverted.} properties of the defect branes.  Starting with a defect brane
whose tension scales as $(g_s)^\alpha$ and using the fact that under S-duality
the string-frame metric occurring in the Nambu-Goto action transforms as
$(g^\prime_{\mu\nu})_S = e^{-8\phi/(D-2)}\,(g_{\mu\nu})_S$ one finds that
under S-duality the value of $\alpha$ changes as
\begin{equation}\label{rule}
\alpha^\prime = -\alpha-4\,.
\end{equation}
This means that under $D$-dimensional S-duality the solitonic defect branes
are mapped to each other while the Dirichlet and Fundamental defect branes are
mapped to defect branes with $\alpha=-3$ and $\alpha=-4$, respectively.  Using
the fact that the number of fundamental branes is well-known this implies that
the number of $\alpha=-4$ branes is known as well.

Applying our supersymmetric brane criterion we find that all Dirichlet defect
branes are supersymmetric (there are no non-supersymmetric Dirichlet branes
within the spinor representation) and the same applies to the charge-conjugate
spinor of defect branes with $\alpha=-3$ \cite{Bergshoeff:2011ee}. On the
other hand, from \cite{Bergshoeff:2011zk} we know that not every component of
the soliton representations corresponds to a supersymmetric brane: ${\rm
  rank}\,T$ solitons out of the ${\rm Adj}_{|T}$ solitons as well as the
singlet soliton are {\sl not} supersymmetric. Using the fact that ${\rm
  rank}\,G={\rm rank}\,T+1$ we therefore conclude that the number $n_{\rm D}$
of supersymmetric defect branes, in each dimension $D\ge 5$, is given by
\begin{equation}\label{nD}
n_{\rm D} = {\rm dim}\, G - {\rm rank}\, G\,,
\end{equation}
in agreement with the statement under Table~\ref{table1}.  The analysis for
$D=3,4$ is the same as the $D\ge 5$ cases because all the fundamental defect
branes within the singlet ($D=4$) and the vector $(D=3$) representations of
T-duality are supersymmetric, and consequently by S-duality the $\alpha=-4$
branes are supersymmetric too, leading again to eq. (\ref{nD}).

\begin{table}[h]
\begin{center}
\begin{tabular}{|c|c||c|c|c|c|c|}
\hline\rule[-1mm]{0mm}{6mm}
$D$&U repr.&$\alpha=0$&$\alpha=-1$&$\alpha=-2$&$\alpha=-3$&$\alpha=-4$\\[.1truecm]
\hline \rule[-1mm]{0mm}{6mm}IIB&$2 \subset {\bf 3}$&&$1$&--&$1$&\\[.05truecm]
\hline \rule[-1mm]{0mm}{6mm}9&$2\subset {\bf 3}_3$&&$1$&--&$1$&\\[.05truecm]
\hline \rule[-1mm]{0mm}{6mm}8&$6\subset {\bf (8,1)}$&&${\bf (2,1)}$&$2\subset {\bf (3,1)}$&${\bf (2,1)}$&\\[.05truecm]
\cline{2-7} \rule[-1mm]{0mm}{6mm}&$2\subset {\bf (1,3)}$&&&$2\subset {\bf (1,3)}$&&\\[.05truecm]
\hline \rule[-1mm]{0mm}{6mm}7&$20\subset {\bf 24}$&&${\bar {\bf 4}}$&$12\subset {\bf 15}$&${\bf 4}$&\\[.05truecm]
\hline \rule[-1mm]{0mm}{6mm}6&$40\subset {\bf 45}$&&${\bf 8}_{\rm V}$&$24\subset {\bf 28}$&${\bf 8}_{\rm V}$&\\[.05truecm]
\hline \rule[-1mm]{0mm}{6mm}5&$72\subset {\bf 78}$&&${\bf 16}$&$40\subset {\bf 45}$&${\overline {\bf 16}}$&\\[.05truecm]
\hline \rule[-1mm]{0mm}{6mm}4&$126\subset {\bf 133}$&${\bf 1}$&${\bf 32}$&$60\subset {\bf 66}$&${\bf 32}$&${\bf 1}$\\[.05truecm]
\hline \rule[-1mm]{0mm}{6mm}3&$240\subset {\bf 248}$&${\bf 14}$&${\bf 64}$&$84\subset {\bf 91}$&${\bf 64}$&${\bf 14}$\\[.05truecm]
\hline
\end{tabular}
\end{center}
 \caption{\sl Defect branes in different dimensions}\label{table2}
\end{table}

Summarizing, we find that in any dimension $5\le D\le 10$ we have a chiral
T-duality spinor of Dirichlet defect branes, a set of solitonic defect branes
that transforms as an anti-symmetric 2-tensor under T-duality and a
charge-conjugate T-duality spinor of $\alpha=-3$ branes, see Table
\ref{table2}. In $D=8$ dimensions the solitonic defect branes split into two
parts: one part that transforms as a positive-dual 2-tensor under the ${\rm
  SO}(2,2)$ T-duality and one part that transforms as a negative-dual
2-tensor. The positive-dual defect branes have a worldvolume vector multiplet
and they transform under U-duality into the defect branes with $\alpha=-1$ and
$\alpha=-3$ which have worldvolume vector multiplets as well. The
negative-dual defect branes have a worldvolume self-dual tensor multiplet and
they transform under U-duality into each other. On top of all these defect
branes we have in $D=4$ dimensions a singlet Fundamental, or $\alpha=0$,
defect brane and in $D=3$ dimensions a T-duality vector of Fundamental defect
branes. These are the usual fundamental string and 0-branes which indeed
become defect branes in $D=4$ and $D=3$ dimensions respectively. Finally, the
$\alpha=-4$ branes corresponding to the S-duals of the $\alpha=0$ branes. This
analysis coincides with the one recently given in \cite{Bergshoeff:2011qk} for
$D\geq 6$ and in \cite{Kleinschmidt:2011vu} for $D\geq 3$.


\section{String and M-theory origin}
\label{sec-stringandmtheoryorigin}

In this Section we wish to consider the string and M-theory origin of the
defect branes of the previous Section, see Table~\ref{table2}.

The string-theory origin of the Fundamental defect branes is the IIA/IIB
Fundamental string supplied with the fundamental wrapping rule
    \begin{eqnarray}\label{Fbranewrapping}
 & & {\rm F} \ \ \ \left\{ \begin{array}{l}
{\rm wrapped} \ \ \ \ \rightarrow\  \ \ {\rm doubled}\\
{\rm unwrapped} \ \ \rightarrow \ \ {\rm undoubled}\quad .
 \end{array} \right.
 \end{eqnarray}
This means that the IIA/IIB fundamental string, upon applying the
wrapping rule~(\ref{Fbranewrapping}) leads to the numbers of
fundamental defect branes given in Table~\ref{table2}. We can
represent the Fundamental string by the NS-NS 2-form field $B_2$ it
couples to, see Table~\ref{table3}.

\begin{table}[h!]
\begin{center}
\begin{tabular}{|c|c|c|c|c|c|c|}
\hline\rule[-1mm]{0mm}{6mm}
$\alpha=0$&\multicolumn{2}{c|}{$\alpha=-1$}
&$\alpha=-2$&\multicolumn{2}{c|}{$\alpha=-3$}
&$\alpha=-4$\\[.1truecm]
\hline\rule[-1mm]{0mm}{6mm}
&IIA&IIB&&IIA&IIB&\\[.05truecm]
\hline\rule[-1mm]{0mm}{6mm}
$B_2$&$C_1$&$C_2$&$D_6$&$E_{8,1}$&$E_8$&$F_{8,6}$\\[.05truecm]
\hline\rule[-1mm]{0mm}{6mm}
&$C_3$&$C_4$&$D_{7,1}$&$E_{8,3}$&$E_{8,2}$&$F_{8,7,1}$\\[.05truecm]
\hline\rule[-1mm]{0mm}{6mm}
&$C_5$&$C_6$&$D_{8,2}$&$E_{8,5}$&$E_{8,4}$&\\[.05truecm]
\hline\rule[-1mm]{0mm}{6mm}
&$C_7$&$C_8$&&$E_{8,7}$&$E_{8,6}$&\\[.05truecm]
\hline
\end{tabular}
\end{center}
\caption{\sl The string-theory origin of all the potentials that couple to
  defect  branes in all dimensions $D\geq 3$: anti-symmetric tensors (coupling
to branes), and  mixed-symmetry fields  (coupling to (generalized) KK
monopoles).  We have not indicated that in $D=3$ also the IIA/IIB pp-wave,
represented by the metric, contributes to the defect 0-branes.}
\label{table3}
\end{table}

Similarly, the string-theory origin of the Dirichlet defect branes given in
Table~\ref{table2} are the IIA and IIB Dirichlet branes supplied with the
Dirichlet wrapping rule
   \begin{eqnarray}\label{Dbranewrapping}
 & & {\rm D} \ \ \ \left\{ \begin{array}{l}
{\rm wrapped} \ \ \ \ \rightarrow\  \ \ {\rm undoubled}\\
{\rm unwrapped} \ \ \rightarrow \ \ {\rm undoubled}\quad .
 \end{array}  \right.
 \end{eqnarray}
The string theory origin of the solitonic defect branes is the
IIA/IIB NS5-brane together with the solitonic wrapping rule
\cite{Bergshoeff:2011ee}
   \begin{eqnarray}\label{Sbranewrapping}
 & & {\rm S} \ \ \ \left\{ \begin{array}{l}
{\rm wrapped} \ \ \ \ \rightarrow\  \ \ {\rm undoubled}\\
{\rm unwrapped} \ \ \rightarrow \ \ {\rm doubled} \quad .
 \end{array} \right.
\end{eqnarray}

To realize the fundamental wrapping rule (\ref{Fbranewrapping}) one needs the
pp-wave. No additional objects (other than the ten-dimensional D-branes
themselves) are needed to realize the Dirichlet wrapping rule
(\ref{Dbranewrapping}). To realize the solitonic wrapping rule
(\ref{Sbranewrapping}) the ten-dimensional Kaluza-Klein (KK) monopole is
needed, but that is not enough: one also needs the so-called {\sl generalized}
KK monopoles. These are extended objects which have, in addition to worldvolume
and transverse directions, {\sl isometry} directions of various kinds with
inequivalent properties. The standard KK monopole only has one isometry
direction.  We find that the string-theory origin of the solitonic defect
branes is given by the NS5-brane, the standard KK monopole and one generalized
KK monopole with two isometry directions.  One can associate mixed-symmetry
fields to generalized KK monopoles, and, in particular, as far as the solitonic
branes are concerned, the standard KK monopole is associated with the field
$D_{7,1}$ and the generalized KK monopole with two isometries with the field
$D_{8,2}$ provided that the restricted reduction rule of
\cite{Bergshoeff:2011mh} is applied. For later convenience we give this
reduction rule below for a mixed-symmetry field $A_{m,n_1 ,n_2 }$
corresponding to a Young tableaux with 3 columns.  \bigskip

\noindent {\bf Restricted reduction rule\,:}\ \ \ for a mixed-symmetry field
$A_{m,n_1 ,n_2 }$ to yield, upon toroidal reduction, a potential corresponding
to a supersymmetric brane, we require that the $n_2$ indices are internal and
along directions parallel to $n_2$ of the $n_1$ indices and $n_2$ of the $m$
indices, and that the remaining $n_1 -n_2$ indices in the second set are also
internal and along directions parallel to $n_1 -n_2$ of the $m$ indices.
\bigskip

To summarize, all solitonic defect branes in any dimensions are generated by the fields
\begin{equation}
D_6\,,\hskip 1truecm D_{7,1}\,,\hskip 1truecm D_{8,2}
\end{equation}
using the restricted reduction rule formulated above.  The field $D_6$ is the
dual of $B_2$, the field $D_{7,1}$ can be seen as a dual graviton at the
linearized level and similarly $D_{8,2}$ is an exotic dual of $B_2$ at the
linearized level.

Using the same reasoning, the string theory origin of the $\alpha=-3$ defect
branes are given by the following branes and generalized KK monopoles:
 \begin{equation}\label{extra}
 E_{8,n}\,,\hskip 1truecm n=0,\dots ,7\, ,
 \end{equation}
 where $n$ is even in the IIB case and odd in the IIA case
 \cite{Bergshoeff:2011ee}.  The eight-form potential $E_8$ (corresponding to
 $n=0$) couples to the S-dual of the D7-brane. The other fields are all exotic
 duals of the RR fields $C_n\,, n=1,\dots 7$ and correspond to generalized KK
 monopole solutions.  In the same way as the D-branes, upon using the
 Dirichlet wrapping rule (\ref{Dbranewrapping}), build up a chiral spinor
 representation of the T-duality group, the S-dual of the D7-brane,
 upon using the exceptional wrapping rule
\begin{eqnarray}\label{Ebranewrapping}
 & & {\rm E} \ \ \ \left\{ \begin{array}{l}
{\rm wrapped} \ \ \ \ \rightarrow\  \ \ {\rm doubled}\\
{\rm unwrapped} \ \ \rightarrow \ \ {\rm doubled}
 \end{array} \right. \quad ,
\end{eqnarray}
builds up the {\sl charge-conjugate} spinor representation of the same
T-duality group. This exceptional wrapping rule is realized through the
generalized monopoles given in (\ref{extra}), using the restricted reduction
rule given above.

Finally, there is no conventional brane origin and corresponding brane
wrapping rule of the $\alpha=-4$ branes.  All these branes follow from the
reduction of generalized KK monopoles. This is to be expected since the only
available $\alpha=-4$ brane in string theory is the S-dual of the D9
brane. However, this is a space-filling brane that upon reduction cannot give
rise to a defect brane. We find that we need two $\alpha =-4$ generalized
monopoles in ten dimensions, that can be associated to the mixed-symmetry
fields
\begin{equation}\label{rule2}
F_{8,6}\,,\hskip 1truecm F_{8,7,1}\,.
\end{equation}
One can easily see that $F_{8,6}$ gives an $\alpha=-4$ singlet 1-brane in four
dimensions, while using the restricted reduction rule in three dimensions one
gets (here we denote with $i$ the internal indices)
\begin{eqnarray}
&& F_{8,6}\ \, \rightarrow F_{1 i_1 ... i_7, i_1 ... i_6}\ \ \ \ (7)\,, \nonumber \\
& & F_{8,7,1} \rightarrow F_{1 i_1...i_7, i_1 ...i_7 , i_1} \ \ (7)
\end{eqnarray}
adding up to a total of 14 0-branes, in agreement with Table \ref{table2}.  A
new feature is that one of the monopoles is described by a mixed-symmetry
field $F_{8,7,1}$ corresponding to a Young tableaux with {\sl three}
columns. This corresponds to a generalized KK monopole with $6+1$ {\sl
  inequivalent} isometry directions.\,\footnote{In which sense they are
  inequivalent, will be discussed later.} The field $F_{8,6}$ can be seen as
an exotic dual of $B_2$, while $F_{8,7,1}$ is an exotic dual of the graviton.
The complete result, including all the fields that after restricted
dimensional reduction give rise to the defect branes, is summarized in Table
\ref{table3}.

One may also consider the M-theory origin of the defect branes. It turns out
that all the fields in Table \ref{table3} have their origin in the
eleven-dimensional fields\,\footnote{We have not indicated the M-theory
  pp-wave which is represented by the metric.}
 \begin{equation}
   A_3\,, \quad A_6\,, \quad A_{8,1}\,, \quad A_{9,3}\,, \quad A_{9,6}\,, \quad A_{9,8,1} \quad . \label{elevendimfieldsextable}
\end{equation}
They correspond to two branes (the M2-and M5-brane), the standard M-theory
monopole and three generalized KK monopoles one of which has {\it two}
inequivalent isometry directions as we will describe in the next Section.  The
fields $A_6$, $A_{9,3}$ and $A_{9,6}$ are duals and exotic duals of the 3-form
potential $A_3$, while $A_{8,1}$ and $A_{9,8,1}$ are duals of the graviton.


\section{Mixed-symmetry fields and monopole solutions}
\label{sec-mixedsymemtry}
In this section we show how the mixed-symmetry fields, together with the
restricted reduction rule, are in one to one correspondence with the
classification of generalized KK monopole solutions of
\cite{LozanoTellechea:2000mc}.\footnote{The relation between mixed-symmetry
  fields and generalized KK monopole solutions has also been recently pointed
  out in \cite{Kleinschmidt:2011vu}.}  We are going to use the following
notation: an extended object of $D$-dimensional string theory with mass
proportional to $g_{s}^{\alpha}$, $T$ transverse dimensions, $p$ spacelike
worldvolume dimensions and $I_{1},I_{2},\dots$ inequivalent isometry
directions,\footnote{In this work we will not have to consider more than two
  inequivalent sets of isometries, but to account for all domain-wall and
  space-filling branes, one has to consider more.} with
$T+p+\sum_{i}I_{i}=D-1$, will be denoted by
$(T,p,I_{1},I_{2},\dots)_{\alpha}$.  We will omit by convention all the
entries to the right of the last non-vanishing $I_{i}$.  Thus, standard
D$p$-branes ($I_{i}=0$) are denoted by $(T,p)_{-1}$, the standard KK monopole
in $D$ dimensions is denoted by $(3,D-5,1)_{-2}$ etc. For M-theory objects
we will omit the subindex
$\alpha$.

The association between the $(p+1)$-form potentials
$B_{2},C_{1},\cdots,C_{8},D_{6},E_{8}$ and $p$-branes is well
established. Mixed-symmetry potentials are associated to generalized KK
monopoles as follows: the symmetry of the potential $A_{m,n}$ if that of a
Young tableau with two columns, one with $m$ rows and one with $n$
rows,\footnote{An anti-symmetric potential is denoted by $A_{m,0} = A_m$.}
and it corresponds to the generalized KK monopole
\begin{eqnarray}\label{rule1}
&&A_{m,n}\hskip 1truecm \leftrightarrow\hskip 1truecm (D-m,m-n-1,n)\, ,
\hskip 1.5truecm{\rm or}\nonumber\\[.5truecm]
&&(T\,, p\,, I)\ \ \leftrightarrow\ \  A_{D-T,I}\, .
\end{eqnarray}
This rule can be extended to include monopoles with two
inequivalent isometry directions as follows
 \begin{eqnarray}\label{rule4}
&&A_{m,n_1,n_2} \ \ \leftrightarrow \ \ (D-m\,, m-n_1-1\,, n_1-n_2\,, n_2)\, ,
\hskip .8truecm {\rm or}\nonumber\\[.5truecm]
&&(T\,, p\,, I_1\,,I_2)\ \ \leftrightarrow\ \  A_{D-T,I_1+I_2,I_2}\,.
\end{eqnarray}

From now on, for simplicity, we will denote the correspondence between the
mixed-symmetry fields and the solutions with an equality, i.e. $A_{m,n} =
(D-m,m-n-1,n)_{\alpha}$.  In this notation, the string-theory origin of the
solitonic defect branes mentioned in the previous Section (the NS5-brane, the
standard KK monopole and one generalized KK monopole) reads
\begin{equation}
D_6 = (4,5)_{-2}\,,\hskip 1truecm D_{7,1} = (3,5,1)_{-2}\,,\hskip 1truecm
D_{8,2} = (2,5,2)_{-2}\, ,
\end{equation}
the string theory origin of the $\alpha=-3$ defect branes reads
 \begin{equation}
\label{extra2}
 E_{8,n} = (2,7-n,n)_{-3}\,,\hskip 1truecm n=0,\dots ,7\, ,
\end{equation}
and the string-theory origin of the $\alpha=-4$ defect branes reads
\begin{equation}
\label{rule2-2}
F_{8,6} = (2,1,6)_{-4}\,,\hskip 2truecm F_{8,7,1} = (2,0,6,1)_{-4}\,.
\end{equation}

\noindent Finally, the M-theory origin of the defect branes reads
\begin{equation}
  \begin{array}{rclrclrcl}
A_3 & = & (6,4)\, ,\hspace{.5cm}&  A_6 & = & (5,5)\, , \hspace{.5cm}&
A_{8,1} & = & (3,6,1)\, , \\
& & & & & & & & \\
A_{9,3}& = & (1,5,3)\hspace{.5cm}&
A_{9,6} & = & (2,2,6)\hspace{.5cm}&  A_{9,8,1} & = & (2,0,7,1)\, .\\
\end{array}
\end{equation}

One advantage of this notation is that it that it makes it easy to write the
mass of a toroidally compactified 10-dimensional monopole solution $(T\,, p\,,
I_1\,, I_2)_{\alpha}$, which is given by ($\ell_{s}=1$)
\begin{equation}
\label{general mass}
M_{(T,p,I_1,I_2)_{\alpha}} = R_1\dots R_p\,(R_{p+1}\dots
R_{p+I_1})^2\,(R_{p+I_1+1}\dots R_{p+I_1+I_2})^3\,(g_s)^\alpha\, ,
\end{equation}
while  for an 11-dimensional monopole it is given by ($\ell_{\rm Planck}^{(11)}/2\pi=1$)
\begin{equation}
\label{general mass2}
M_{(T,p,I_1,I_2)} = R_1\dots R_p\,(R_{p+1}\dots
R_{p+I_1})^2\,(R_{p+I_1+1}\dots R_{p+I_1+I_2})^3\, .
\end{equation}
Here the $R's$ are the compactification radii in the spacelike worldvolume and
two isometry directions. It is this different dependence on the
compactification radii that makes the isometry directions inequivalent. For
instance, the mass of the $F_{8,7,1} = (2,0,6,1)_{-4}$ generalized KK monopole
is given by
\begin{equation}\label{mass}
M_{(2,0,6,1)_{-4}} = (R_1\dots R_6)^2\, (R_7)^3\, (g_s)^{-4}\,,
\end{equation}
where $1,\dots ,7$ indicate the $6+1$ isometry directions.  Similarly, the
mass of the $A_{9,8,1}=(2,0,7,1)$ generalized KK monopole is given by
\begin{equation}\label{mass2}
M_{(2,0,7,1)} = (R_1\dots R_7)^2 (R_8)^3\,,
\end{equation}
where $1,\dots ,8$ refer to the $7+1$ isometry directions.

This identification is based on the consistency between the restricted
reduction rules of the potentials and the dimensional reduction of the
objects.  One can reduce a monopole solution given by $(T\,, p\,, I_1\,, I_2)$
in four different ways: over a transverse ($T$), worldvolume ($p$) or one of
the two inequivalent isometry directions ($I_{1},I_{2}$). This leads to brane
solutions as soon as one has reduced over all isometry directions.  The branes
corresponding to such solutions couple to a number of potentials. In order to
obtain the same number of potentials following from the reduction of the
mixed-symmetry fields one must use the restricted reduction rule formulated in
Section 3.

As an example we consider the string and M-theory origin of the eight $D=8$
defect brane solutions, see Table~\ref{table2}.  We have indicated the IIA and
IIB string theory origin of these eight solutions in
Table~\ref{table5}. Assuming that we reduce over the $i=6,7$ directions the
three eight-dimensional dilatons are given by $g_s\,,R_6$ and $R_7$, where
$R_6$ and $R_7$ are the radii in the $6$ and $7$ directions. The IIA origin of
the remaining 4 axions is given by $g_{67}\,, B_{67}\,, C_6$ and $C_7$ where
$C_\mu$ is the RR vector. Similarly, the IIB origin of the same axions is
given by $g_{67}\,, B_{67}\,, C_{67}$ and $C_0$ where $C_0$ is the IIB
axion. The two transverse directions of the defect brane are 8 and 9.

\begin{table}[h!]
\begin{center}
\begin{tabular}{|c|c|c|}
\hline\rule[-1mm]{0mm}{6mm}
IIA&monopole&$M={\rm mass}/V_5$\\[.05truecm]
\hline\rule[-1mm]{0mm}{6mm}
$\alpha=-1$&$C_7= (3,6)_{-1}$&$M=R_i\,(g_s)^{-1}$ \\[.05truecm]
\hline\rule[-1mm]{0mm}{6mm}
$\alpha=-2$&$D_6=(4,5)_{-2}$&$M=(g_s)^{-2}$\\[.05truecm]
\hline\rule[-1mm]{0mm}{6mm}
&$D_{7,1}=(3,5,1)_{-2}$&$M=(R_i)^2\,(g_s)^{-2}$\\[.05truecm]
\hline\rule[-1mm]{0mm}{6mm}
&$D_{8,2}=(2,5,2)_{-2}$&$M=(R_6R_7)^2\,(g_s)^{-2}$\\[.05truecm]
\hline\rule[-1mm]{0mm}{6mm}
$\alpha=-3$&$E_{8,1}=(2,6,1)_{-3}$&$M=R_i\,(R_{i+1})^2\,(g_s)^{-3}$\\[.05truecm]
\hline
&&\\[.05truecm]
\hline\rule[-1mm]{0mm}{6mm}
IIB&monopole&$M={\rm mass}/V_5$\\[.05truecm]
\hline\rule[-1mm]{0mm}{6mm}
$\alpha=-1$&$C_8=(2,7)_{-1}$&$M=R_6R_7\,(g_s)^{-1}$\\[.05truecm]
\hline\rule[-1mm]{0mm}{6mm}
&$C_6=(4,5)_{-1}$&$M=(g_s)^{-1}$\\[.05truecm]
\hline\rule[-1mm]{0mm}{6mm}
$\alpha=-2$&$D_6=(4,5)_{-2}$&$M=(g_s)^{-2}$\\[.05truecm]
\hline\rule[-1mm]{0mm}{6mm}
&$D_{7,1}=(3,5,1)_{-2}$&$M=(R_i)^2\,(g_s)^{-2}$\\[.05truecm]
\hline\rule[-1mm]{0mm}{6mm}
&$D_{8,2}=(2,5,2)_{-2}$&$M=(R_6R_7)^2\,(g_s)^{-2}$\\[.05truecm]
\hline\rule[-1mm]{0mm}{6mm}
$\alpha=-3$&$E_8=(2,7)_{-3}$&$M=R_6R_7\,(g_s)^{-3}$\\[.05truecm]
\hline\rule[-1mm]{0mm}{6mm}
&$E_{8,2}=(2,5,2)_{-3}$&$M=(R_6R_7)^2\,(g_s)^{-3}$\\[.05truecm]
\hline
\end{tabular}
\end{center}
 \caption{\sl This table indicates the string theory origin of the
eight $D=8$  half-supersymmetric defect brane solutions. The common
factor $V_5$ in the expression for the mass is given by $V_5 = R_1\dots
R_5$. We have set $\ell_s=1$. The 8,9 directions are the two transverse
directions.  The free index $i=6,7$ indicates two defect brane solutions.
In the IIA case, the $D_6$ and $D_{8,2}$ solutions lead to the two tensor defect branes, while in the IIB case they arise from the $D_{7,1}$ solution.
}\label{table5}
\end{table}

The IIA/IIB string theory origin of all eight $D=8$ supersymmetric defect
branes are given in Table~\ref{table5}. Note that each object has a different
mass.  All these objects and corresponding solutions are known in the
literature. For instance, in the IIA case, the $(3,6)_{-1}$ object is the
D6-brane. This object gives rise to two defect branes depending on whether we
take $i=6$ or $i=7$ along the worldvolume directions of the
D6-brane. $(4,5)_{-2}$ is the NS5A-brane and $(3,5,1)_{-2}$ is the standard
KK5A monopole. $(2,5,2)_{-2}$ is a generalized KK monopole whose M-theory
origin is another generalized KK monopole: $A_{9,3}=(2,5,3)$.  The
corresponding explicit solution of the latter can be found in eq.~(3.9) of
\cite{LozanoTellechea:2000mc}.\,\footnote{Some of the generalized monopoles
  have been constructed using ${\rm E}_{11}$ techniques
  \cite{Englert:2007qb}.}  Finally, $(2,6,1)_{-3}$ is the reduction of the
$(3,6,1)$ standard M-theory monopole solution over one of its transverse
directions and corresponds to the $p=6$ case of eq.~(1.1) of
\cite{LozanoTellechea:2000mc}. Together, the M-theory origin of all the IIA
solutions is given by the $(5,5)$ M5-brane solution, the $(3,6,1)$ standard
M-theory KK monopole and the $(2,5,3)$ generalized KK monopole solution.

In the IIB case the two $\alpha=-1$ objects are the D5-brane and the
D7-brane. The three $\alpha=-2$ objects are the same as in the IIA
case. Finally, the two $\alpha=-3$ objects are the S-dual of the D7-brane and
the $(2,5,2)_{-3}$ generalized KK monopole.

The masses of all the objects of Table \ref{table5} transform into each other
under the T-duality rules
\begin{equation}
R\ \ \rightarrow \ \ 1/R\,,\hskip 1truecm g_s\ \ \rightarrow \ \ g_s/R\,,
\end{equation}
in agreement with the T-duality representations given in Table
\ref{table2}. Note that the mass of the solutions is not left invariant under
this T-duality. The mass multiplets form representations of the ${\rm SO}(3)$
subgroup of the ${\rm SO}(2,2)$ T-duality group. Note that under the S-duality
rules
\begin{equation}\label{Srules}
g_s\ \ \rightarrow 1/g_s\,,\hskip 1truecm R\  \ \rightarrow R/(g_s)^{1/2}
\end{equation}
the masses do not transform in agreement with the S-duality rule
(\ref{rule}). This is because the S-duality underlying (\ref{rule})
refers to the {\sl eight-dimensional} dilaton whereas the S-duality
rules (\ref{Srules}) refer to the {\sl ten-dimensional} dilaton.

\section{Duality relations and explicit solutions}
\label{sec-explicitsolutions}

In this section we want to show that the linearized duality relations that the
mixed-symmetry fields satisfy can be used to deduce the behaviour of the
fields of the corresponding solution. We consider as a first example the
ten-dimensional field $B_2$ together with all its generalized duals $D_6$,
$D_{8,2}$ and $F_{8,6}$.  For each of these fields, there is a $(T,p,I)$
solution in which the field can be considered to be electric, that is with
non-zero components along the $p+1$ worldvolume directions and along the
isometry directions.  We now want to show that using linearized duality
relations each of these solutions becomes a solution in which only the $B_2$
field occurs. The duality relation reveals in each case the particular form
that the $B_2$ field takes.

We denote the $T$ transverse directions with $\omega^a$, with $a=1,...,T$, the
worldvolume direction with $y^\mu = (t,y^1,...,y^p)$ and the isometry
directions with $z^m$, with $m=1,...,I$. In all cases the fields only depend
on the transverse directions $\omega$.  We start considering the $B_2$
solution. This is the solution $(8,1)_0$, and simply corresponds to an
electric field $B_{\mu\nu} (\omega)$. We next consider the $D_6$ solution
$(4,5)_{-2}$. This corresponds to a non-vanishing $D_{\mu_1...\mu_6}
(\omega)$. Using the duality relation

  \begin{equation}
 \partial_{a} D_{\mu_1...\mu_6} \sim \epsilon_{a \mu_1 ...\mu_6}{}^{ b_1 b_2
   b_3}
\partial_{b_1} B_{b_2 b_3}
\end{equation}
we see that this corresponds to a $B_2$ field along the four transverse directions
$B_{a_1 a_2}(\omega)$.

We next consider the field $D_{8,2}$. The solution $(2,5,2)_{-2}$ has two
isometries, and corresponds to turning on the components $D_{\mu_1 ...\mu_6
  mn,mn} (\omega)$. Dualizing we get\,\footnote{Note that the duality
  relations involving mixed-symmetry fields we use in this section are not
  truly ten-dimensional ones. They are only applied to solutions that exhibit
  a number of isometries.  Effectively, this means that, after reduction over
  the isometry directions, we apply standard lower-dimensional duality
  relations between forms. We thank Axel Kleinschmidt for a discussion on this
  point.}
  \begin{equation}
  \partial_a \partial_b D_{\mu_1 ...\mu_6 mn,mn} \sim \epsilon_{a \mu_1 ...\mu_6 mn}{}^c
 \partial_b \partial_c B_{mn}
\end{equation}
which means that the solution can be seen as a solution in which one turns on
$B_2$ along the isometry directions, $B_{mn}(\omega)$. The linearized duality
relation is at second order in derivatives because the field has
mixed-symmetry with two sets of antisymmetric indices
\cite{Curtright:1980yk,Hull:2001iu}.

We finally consider the field $F_{8,6}$ corresponding to the solution
$(2,1,6)_{-4}$ with six isometries. This solution is carried by the electric
mixed-symmetry field $F_{\mu_1 \mu_2 m_1 ...m_6, m_1 ...m_6} (\omega)$ which
can be dualized as follows:
   \begin{equation}
  \partial_a \partial_b F_{\mu_1 \mu_2 m_1 ...m_6,m_1 ...m_6} \sim \epsilon_{a \mu_1 \mu_2 m_1 ...m_6 }{}^c \epsilon_{b m_1 ...m_6 }{}^{d \nu_1 \nu_2} \partial_c \partial_d B_{\nu_1 \nu_2}\,.
\end{equation}
This corresponds to a solution with $B_{\mu\nu}(\omega)$ non-vanishing,
exactly as in the first case, but now, since there are isometry directions,
this $B_2$ is not electric.

The same reasoning can be applied to the solutions $(2,7-n,n)_{-3}$
corresponding to the fields $E_{8,n}$.  These solutions have $n$
isometries. The field $E_{\mu_1 ...\mu_{8-n} m_1 ...m_n, m_1 ...m_n}(\omega)$
is non-vanishing, and can be dualized according to
   \begin{eqnarray}
  & & \partial_a \partial_b E_{\mu_1 ... \mu_{8-n} m_1 ...m_n,m_1 ...m_n} \sim \epsilon_{a \mu_1 ... \mu_{8-n}  m_1 ...m_n }{}^c \partial_b \partial_c   C_{m_1 ...m_n} \nonumber \\
& & \sim  \epsilon_{a \mu_1 ... \mu_{8-n}  m_1 ...m_n }{}^c  \epsilon_{b m_1 ...m_n}{}^{\nu_1 ...\nu_{8-n} d} \partial_c \partial_d C_{\nu_1 ...\nu_{8-n}} \quad ,
\end{eqnarray}
corresponding to a solution with the RR field $C_n$ along the isometry
directions (or a dual RR field $C_{8-n}$ along the worldvolume directions).

We now consider the purely gravitational solutions. The KK monopole solution
is $(3,5,1)_{-2}$. The corresponding ten-dimensional field is $D_{7,1}$, and
turning on the component $D_{\mu_1 ...\mu_6 m,m}(x)$ the linearized duality
relation becomes
  \begin{equation}
\partial_a \partial_b D_{\mu_1 ...\mu_6 m,m} \sim \epsilon_{a \mu_1 ...\mu_6 m}{}^{cd} \partial_c \partial_b h_{dm} \quad ,
\end{equation}
corresponding to a linearized graviton fluctuation of the form
  \begin{equation}
h_{a m}(x) \quad .
\end{equation}
This is the well-known KK monopole solution.

The other (generalized) KK monopole solution is $(2,0,6,1)_{-4}$ where now
there are two sets of isometries: a six-plet and a singlet isometry
direction. The corresponding field is $F_{8,7,1}$ which has non-vanishing
components $F_{\mu m_1 ...m_7 , m_1 ... m_7, m_1}$. This corresponds to a
linearized graviton given by the duality relation
  \begin{equation}
  \partial_a \partial_b \partial_c F_{\mu m_1 ...m_7, m_1 ...m_7 , m_1} \sim \epsilon_{a \mu m_1 ...m_7}{}^d \epsilon_{b m_1 ...m_7}{}^{\nu e} \partial_d \partial_e \partial_c h_{\nu m_1} \quad ,
\end{equation}
which corresponds to a linearized graviton fluctuation
  \begin{equation}
h_{\mu m} (x)
 \end{equation}
where $m$ is the singlet isometry direction, and thus this solution is a pp-wave.

Finally, we consider the eleven-dimensional solutions. Repeating the
a\-na\-lysis just done for the $B_2$ field and its duals in ten dimensions,
one can deduce that the solution $A_{9,3} = (2,5,3)$ corresponds to the field
$A_3$ along the isometry directions, while the solution $A_{9,6}= (2,2,6)$
corresponds to $A_3$ along the worldvolume directions. The gravitational
solutions $A_{8,1} = ( 3,6,1)$ and $A_{9,8,1} = ( 2,0,7,1)$ are exactly as in
the ten-dimensional case.

These results can be tested by looking into the explicit supergravity
solutions given in \cite{LozanoTellechea:2000mc}, which we reproduce here for
the sake of completeness. Recently, some of these solutions have been
rederived in \cite{Kleinschmidt:2011vu} by performing U-duality
transformations on known solutions in the ${\rm E}_{11}$ framework.

Let us start with the 10-dimensional (string-theory) fields in
Table~\ref{table3}: $B_{2}=(8,1)_{0}$ is the Fundamental (IIA/IIB) string,
$C_{p+1}=(9-p,p)_{-1}$ are the Dirichlet $p$-branes, $D_{6}=(4,5)_{-2}$ is the
(IIA/IIB) NS5 brane, $D_{7,1}=(3,6,1)_{-2}$ is the standard (IIA/IIB) KK
monopole. The explicit form of all these solutions is well known.

The solution corresponding to $D_{8,2}= (2,5,2)_{-2}$ is, in the string
frame\footnote{In all these defect brane solutions function ${\cal H}={\cal
    H}(\omega)= A+iH$ is a complex, holomorphic, (multivalued) function of
  ${\omega}$.}
\begin{equation}
\label{eq:p=5Sdual}
\begin{array}{rcl}
ds_{s}^2 & = &
dt^{2}-d\vec{y}_{5}^{\ 2} -Hd\omega d\bar{\omega}
-\displaystyle{\frac{H}{{\cal H}\bar{\cal H}}}
d\vec{z}_{2}^{\ 2}\, ,\\
& & \\
e^{\phi}  & = &
 \left( \displaystyle{\frac{H}{{\cal H}\bar{\cal H}}}
\right)^{\frac{1}{2}} \, ,\\
& & \\
B_{(6)\, ty^{1}\cdots y^{5}} & =  &
\left(\displaystyle{ \frac{H}{{\cal H}\bar{\cal H}} }\right)^{-1}\, ,
\hspace{.5cm}
B_{(2)\, z^{1}z^{2}} =
-\displaystyle{ \frac{A}{{\cal H}\bar{\cal H}} }\, ,
\end{array}
\end{equation}
which is eq.~(3.1) of \cite{LozanoTellechea:2000mc}.  Observe that, as anticipated,
$B_{2}$ only has non-vanishing components in the two isometric directions.  The
solutions corresponding to the $E_{8,n}=(2,7-n,n)_{-3}$ are, with $7-n=p$ and
in the string frame, given by
\begin{equation}
\label{eq:typeIIsolutions}
\begin{array}{rcl}
ds_{s}^{2} & = &
\left(\displaystyle{\frac{H}{{\cal H}\bar{\cal H}}}\right)^{-1/2}
\left[dt^{2}-d\vec{y}_{p}^{\ 2}
-Hd\omega d\bar{\omega}\right]
-\left(\displaystyle{\frac{H}{{\cal H}\bar{\cal H}}}\right)^{1/2}
d\vec{z}_{7-p}^{\ 2}\, ,\\
& & \\
e^{\phi} & = &
 \left( \displaystyle{\frac{H}{{\cal H}\bar{\cal H}}}
\right)^{\frac{3-p}{4}} \, ,\\
& & \\
C_{(p+1)\, ty^{1}\cdots y^{p}} & = &
(-1)^{\left[\frac{(p+1)}{2}\right]}
\left( \displaystyle{\frac{H}{{\cal H}\bar{\cal H}}}
\right)^{-1}\, ,
\hspace{.5cm}
C_{(7-p)\, z^{1}\ldots z^{7-p}}
=  -\displaystyle{ \frac{A}{{\cal H}\bar{{\cal H}}} }\, ,
\end{array}
\end{equation}
which is eq.~(1.1) of \cite{LozanoTellechea:2000mc}. As anticipated, this
corresponds to the field $C_{8-n}$ along the worldvolume directions or the
dual field $C_n$ along the isometry directions. The solution corresponding to
the $F_{8,6}= (2,1,6)_{-4}$ is
\begin{equation}
\label{eq:p=1Sdual}
\begin{array}{rcl}
ds_{s}^2 & = &
\left(\displaystyle{\frac{H}{{\cal H}\bar{\cal H}}}\right)^{-1}
\left[dt^{2}-dy^{2} -Hd\omega d\bar{\omega}\right]
-d\vec{z}_{6}^{\, 2}\, ,\\
& & \\
e^{\phi} & = &
 \left( \displaystyle{\frac{H}{{\cal H}\bar{\cal H}}}
\right)^{-\frac{1}{2}} \, ,\\
& & \\
B_{(2)\, ty} & = &
-\left(\displaystyle{ \frac{H}{{\cal H}\bar{\cal H}} }\right)^{-1}\, ,
\hspace{.5cm}
B_{(6)\, z^{1}\cdots z^{6}} \,=\,
\displaystyle{ \frac{A}{{\cal H}\bar{\cal H}} }\, , \\
\end{array}
\end{equation}
which is eq.~(3.2) of \cite{LozanoTellechea:2000mc} and as anticipated
corresponds to the field $B_2$ along the worldvolume directions.  The purely
gravitational solution corresponding to $F_{8,7,1}=(2,0,6,1)_{-4}$ is
\begin{equation}
ds^{2}=-2dtdy - \frac{H}{{\cal H}\bar{\cal H}} dy^{2}
-{\cal H}\bar{\cal H}d\omega d\bar{\omega} -d\vec{z}^{\ 2}_{6}\, ,
\end{equation}
which is eq.~(3.11) of \cite{LozanoTellechea:2000mc} and is  a $pp$-wave.

As for the explicit solutions corresponding to the eleven-dimensional fields
in eq.~(\ref{elevendimfieldsextable}), $A_{3}= (8,2)$ and $A_{6}=(5,5)$ are
the M2 and M5 branes, $A_{8,1}=(3,6,1)$ is the standard KK monopole and all
their solutions are well known. $A_{9,3}= (2,5,3)$ is given by
\begin{equation}
\begin{array}{rcl}
ds^{2} & = &
\left( \displaystyle{\frac{H}{{\cal H}\bar{\cal H}}}\right)^{-1/3}
\left[dt^{2}-d\vec{y}^{\ 2}_{5}-H d\omega d\bar{\omega}\right]
-\left( \displaystyle{\frac{H}{{\cal H}\bar{\cal H}}}\right)^{2/3}
d\vec{z}^{\ 2}_{3}\, ,\\
& & \\
A_{(6)\, ty^{1}\cdots y^{5}}  & =  &
-\left( \displaystyle{\frac{H}{{\cal H}\bar{\cal H}}}\right)^{-1}\, ,
\hspace{1cm}
A_{(3)\, z^{1}z^{2}z^{3}}  =
-\displaystyle{\frac{A}{{\cal H}\bar{\cal H}}}\, ,\\
\end{array}
\end{equation}
which is eq.~(3.9) of \cite{LozanoTellechea:2000mc} and as anticipated has
$A_3$ along the isometry directions, while $A_{9,6}= (2,2,6)$ is given in
\begin{equation}
\begin{array}{rcl}
ds^{2} & = &
\left( \displaystyle{\frac{H}{{\cal H}\bar{\cal H}}}\right)^{-2/3}
\left[dt^{2}-d\vec{y}^{\ 2}_{2} -Hd\omega d\bar{\omega}\right]
-\left( \displaystyle{\frac{H}{{\cal H}\bar{\cal H}}}\right)^{1/3}
d\vec{z}^{\ 2}_{6}\, ,\\
& & \\
A_{(3)\, ty^{1}y^{2}} & = &
-\left( \displaystyle{\frac{H}{{\cal H}\bar{\cal H}}}\right)^{-1}\, ,
\hspace{1cm}
A _{(6)\, z^{1}\cdots z^{6}} =
\displaystyle{\frac{A}{{\cal H}\bar{\cal H}}}\, ,\\
\end{array}
\end{equation}
which is eq.~(3.8) of \cite{LozanoTellechea:2000mc} and as anticipated has
$A_3$ along the worldvolume directions. Finally, $A_{9,8,1}= (2,0,7,1)$ is
given by the purely gravitational solution
\begin{equation}
ds^{2}=-2dtdy - \frac{H}{{\cal H}\bar{\cal H}} dy^{2}
-{\cal H}\bar{\cal H}d\omega d\bar{\omega} -d\vec{z}^{\ 2}_{7}\, ,
\end{equation}
which is eq.~(3.7) of \cite{LozanoTellechea:2000mc} and again corresponds to a pp-wave.

\begin{table}[ht!]
\begin{small}
\begin{center}
\begin{tabular}{|c|c|c|c|c|c|c|c|}
\hline
$D$&$R$&$p=0$&$p=1$&$p=2$&$p=3$&$p=4$&$p=5$\\[.1truecm]
\hline \rule[-1mm]{0mm}{6mm} IIA&{\bf 1}&{\bf 1}&{\bf 1}&{\bf 1}&&{\bf 1}&${\bf 1}$\\[.05truecm]
\cline{3-8} \rule[-1mm]{0mm}{6mm}&&D0&F1&D2&&D4&${\rm S}^\prime$5\ +\\[.05truecm]
\cline{3-7}  \rule[-1mm]{0mm}{6mm}&&&--&&&D6&KK5\\[.05truecm]
\hline
\hline \rule[-1mm]{0mm}{6mm} IIB&SO(2)&&{\bf 2}&&{\bf 1}&&${\bf 1}^+ + {\bf 2}^+$\\[.05truecm]
\cline{3-8} \rule[-1mm]{0mm}{6mm}&&&F1+D1&&D3&&${\rm KK}^\prime$5 +\\[.05truecm]
\cline{3-7} \rule[-1mm]{0mm}{6mm} &&&&&&&(D5+S5)\\[.05truecm]
\hline
\hline \rule[-1mm]{0mm}{6mm}
9&SO(2)&${\bf 1}+{\bf 2}$&{\bf 2}&{\bf 1}&{\bf 1}&${\bf 1}+{\bf 2}$&\\[.05truecm]
\cline{2-7} \rule[-1mm]{0mm}{6mm}&&F0\,+&F1+D1&D2&D3&KK4\,+&\\[.05truecm]
\cline{4-6} \rule[-1mm]{0mm}{6mm}&&(F0+D0)&&&&(D4+S4)&\\[.05truecm]
\cline{3-7} \rule[-1mm]{0mm}{6mm}&&&&&&${\rm S}^\prime$5\,+&\\[.05truecm]
\cline{3-6} \rule[-1mm]{0mm}{6mm}&&&&&&(D5+S5)&\\[.05truecm]
\hline
\hline \rule[-1mm]{0mm}{6mm} 8&U(2)& $ 2 \times {\bf 3}$ &{\bf 3} &
$2 \times {\bf 1}$& ${\bf 1} + {\bf 3}$
&${\bf 3}^+ + {\bf 3}^-$&\\[.05truecm]
\cline{2-7} \rule[-1mm]{0mm}{6mm}&&2$\times$(2F0+D0)&F1+2D1&2$\times$D2&KK3\ +&&\\[.05truecm]
\rule[-1mm]{0mm}{6mm}&&&&&(2D3+S3)&(D4+2S4)\ +&\\[.05truecm]
\cline{3-6} \rule[-1mm]{0mm}{6mm}&&&&&&(D4+2S4)&\\[.05truecm]
\hline
\hline \rule[-1mm]{0mm}{6mm} 7&Sp(4)& {\bf 10} & {\bf 5} & ${\bf 1} +{\bf 5}$ &{\bf 10} &&\\[.05truecm]
\cline{2-6} \rule[-1mm]{0mm}{6mm}&&6F0+4D0&F1+4D1&KK2\ +&4D3+6S3&&\\[.05truecm]
\rule[-1mm]{0mm}{6mm}&&&&(4D2+S2)&&&\\[.05truecm]
\hline
\end{tabular}
\end{center}
\end{small}
 \caption{\sl This table indicates the $R$-representations of the
  $p$-form central charges and the corresponding  standard supersymmetric
p-branes of $7\le D\le 10$ maximal supergravity.  A prime indicates that the
worldvolume multiplet is not a vector but a  tensor multiplet. The pp-wave
corresponds to the translation generator.}
  \label{table8}
\end{table}


\section{Central charges}
\label{sec-centralcharges}

It is well-known that there is a 1-1 correspondence between standard branes
and the central charges in the supersymmetry algebra in type~II string- and
M-theory. The standard branes, for $3\le D\le 10$ dimensions have a universal
behaviour with respect to T-duality. For each dimension they are given by a
singlet and vector of Fundamental branes, a chiral spinor of D-branes and
anti-symmetric tensors of solitonic branes. On top of this we have in each
dimension a pp-wave and a $(3,D-5, 1)_{-2}$ standard KK monopole. The pp-wave is
represented by the translation generator whereas all other branes are
represented by the most general central charges in the supersymmetry
algebra. This is summarized in  Tables
\ref{table8} and \ref{table9}, that indicate the $R$-representations of
the $p$-form central charges and the corresponding standard supersymmetric
p-branes of $3\le D\le 10$ maximal supergravity. In the Tables, if applicable, we have
indicated the space-time duality of the central charges with a superscript
$\pm$. We also use the following abbreviations in the Tables: F (Fundamental), D
(D-brane), S (Soliton), and KK (Kaluza-Klein Monopole).  All branes have
worldvolume vector multiplets except for the ones indicated by a prime.  Note
that in $D=3$ dimensions there are no standard branes.

\begin{table}[h]
\begin{center}
\begin{tabular}{|c|c|c|c|c|c|}
\hline
$D$&$R$&$p=0$&$p=1$&$p=2$&$p=3$\\[.1truecm]
\hline \rule[-1mm]{0mm}{6mm} 6&Sp(4)$\times$&$({\bf 4},{\bf 4})$&$({\bf 1},{\bf 1})$ & $({\bf 4},{\bf 4})$& $({\bf 10},{\bf 1})^+$ \\[.05truecm]
\rule[-1mm]{0mm}{6mm}&Sp(4)  & &$({\bf 1},
{\bf 5})$ & &$({\bf 1},{\bf 10})^-$ \\[.05truecm]
\rule[-1mm]{0mm}{6mm} & & &$({\bf 5},
{\bf 1})$ & &\\[.05truecm]
\cline{2-6}\rule[-1mm]{0mm}{6mm}&&8F0+8D0&KK1&8D2+8S2&\\[.05truecm]
\rule[-1mm]{0mm}{6mm}&&&F1+4D1&&\\[.05truecm]
\rule[-1mm]{0mm}{6mm}&&&4D1+S1&&\\[.05truecm]
\hline
\hline \rule[-1mm]{0mm}{6mm} 5&Sp(8)& {\bf 1} + {\bf 27}& {\bf 27}& {\bf 36}&\\[.05truecm]
\cline{2-5}\rule[-1mm]{0mm}{6mm}&&KK0\ +&F1+16D1&&\\[.05truecm]
\rule[-1mm]{0mm}{6mm}&&10F0+16D0+S0&+10S1&&\\[.05truecm]
\hline
\hline \rule[-1mm]{0mm}{6mm} 4&SU(8)& ${\bf 28} + {\bf \overline{28}}$ & {\bf 63} & ${\bf 36}^+ + {\bf \overline{36}}^-$ &\\[.05truecm]
\cline{2-5}\rule[-1mm]{0mm}{6mm}&&12F0+16D0&&&\\[.05truecm]
\rule[-1mm]{0mm}{6mm}&&16D0+12S0&&&\\[.05truecm]
\hline
\hline \rule[-1mm]{0mm}{6mm} 3&SO(16)& {\bf 120} & {\bf 135} &&\\[.05truecm]
\hline
\end{tabular}
\end{center}
 \caption{\sl This table indicates the $R$-representations of the
  $p$-form central charges and the corresponding  standard supersymmetric p-branes of $3\le D\le 6$ maximal supergravity.
  The pp-wave corresponds to the translation generator. }
  \label{table9}
\end{table}

\begin{table}[h]
\begin{center}
\begin{tabular}{|c|c|c|c|c|c|c|}
\hline
$D$&$H$&$n=0$&$n=1$&$n=2$&$n=3$&$n_{\rm D}$\\[.1truecm]
\hline \rule[-1mm]{0mm}{6mm} IIB &SO(2)&&&&{\bf 1}&{2} \\[.05truecm]
\hline \rule[-1mm]{0mm}{6mm} 9&SO(2)&&&&{\bf 1}&{2} \\[.05truecm]
\hline \rule[-1mm]{0mm}{6mm} 8&U(2)&&&&{\bf 3+1}&{6} + ${ 2} $\\[.05truecm]
\hline \rule[-1mm]{0mm}{6mm} 7&Sp(4)&&&&{\bf 10}&{20} \\[.05truecm]
\hline \rule[-1mm]{0mm}{6mm} 6&Sp(4) $\times $ Sp(4)&&&&$({\bf 10,1})^+ +({\bf 1,10})^-$& {40}\\[.05truecm]
\hline \rule[-1mm]{0mm}{6mm} 5&Sp(8)&&&{\bf 36}&&{72} \\[.05truecm]
\hline \rule[-1mm]{0mm}{6mm} 4&SU(8)&&{\bf 63}&&&{ 126} \\[.05truecm]
\hline \rule[-1mm]{0mm}{6mm} 3&SO(16)&{\bf 120}&&&&{ 240} \\[.05truecm]
\hline
\end{tabular}
\end{center}
 \caption{\sl The number $n_{{\rm D}}$ of supersymmetric defect branes
is  twice the number $n_{{\rm Z}}$ of corresponding n-form central charges.}
\label{table6}
\end{table}

One does not expect a similar 1-1 relation to hold between the non-standard
branes and the central charges of the supersymmetry algebra. The reason is
that these non-standard branes are not asymptotically flat and therefore the
standard Poincar\'e supersymmetry algebra is not realized at spatial
infinity. Nevertheless, since we calculated the number $n_{\rm D}$ of
supersymmetric defect branes, it is of interest to compare these numbers with
the number $n_{\rm Z}$ of relevant $p$-form central charges.\,\footnote{We do
  not consider here the charges corresponding to the generalized KK
  monopoles.} These are the 3-form central charges for $D\ge 6$ and the
$(D-3)$-form central charges for for $3\le D \le 5$. We have collected these
numbers in Table \ref{table6}. We observe that there is a universal 2-1
relation between $n_D$ and $n_Z$, i.e.~we find that $n_{{\rm D}}=2n_{{\rm
    Z}}$. The reason that this is the case is due to the universal behaviour
of the central charges and defect branes. In any dimension the central charges
corresponding to defect branes transform in the adjoint representation of the
$R$-symmetry group $H$, which is the maximal compact subgroup of the U-duality
group $G$, i.e.~we always have that $n_{{\rm Z}} = {\rm dim}\, H$.  On the
other hand, we found that the number of supersymmetric defect branes $n_{\rm
  D}$ is universally given by $n_{\rm D} = {\rm dim}\,G - {\rm rank}\,G$.  We
now use that the U-duality groups of all maximal supergravity theories are of
split-form and therefore we have that ${\rm dim}\, H = P$ and ${\rm dim}\,G -
{\rm rank}\,G =2P$ where $P$ is the number of positive roots. This indeed
implies that $n_{\rm D}=2n_{\rm Z}$.

\section{Conclusions}
\label{sec-conclusions}

In this work we have discussed some basic properties of branes with
co-di\-men\-sion 2, i.e.~defect branes. Requiring the existence of a
supersymmetric gauge-invariant WZ term we gave a full classification of these
branes, see Table~\ref{table2}. Their string and M-theory origin as
seven-branes and a set of generalized KK monopoles was determined. These
included monopoles with two inequivalent isometry directions. We explained why
the number $n_{\rm D}$ of supersymmetric defect branes does not equal the
number $n_{\rm P}$ of $(D-2)$-form potentials or the number $n_{\rm S}$ of
coset scalars and we presented the string and M-theory origin of all defect
branes. As an example we gave explicit results for the $D=8$ case. We observed
that the number $n_{\rm D}$ of supersymmetric defect branes is always twice
the number $n_{\rm Z}$ of central charges in the supersymmetry algebra and we
explained why this is the case.

There is a simple alternative way to count the number of supersymmetric defect
branes and to verify that for a U-duality group $G$ the number $n_{\rm D}$ of
supersymmetric defect branes is given by $n_{\rm D}= {\rm dim}\, G - {\rm
  rank}\, G$.  Each basic half-supersymmetric defect brane is carried by an
axion-dilaton combination that parametrizes an ${\rm SL}(2,\mathbb{R})$
subgroup of the U-duality group. Together with the S-dual defect brane this
leads to two branes for each inequivalent embedding of ${\rm
  SL}(2,\mathbb{R})$ into $G$. For instance, for $G={\rm SL}(n,\mathbb{R})$,
which is the case for $D=7$ and $D=9$, one has to choose 2 out of the $n$
directions. This leads to $n(n-1)/2$ inequivalent embeddings and hence
$n(n-1)$ supersymmetric defect branes. On the other hand, for $G={\rm
  SL}(n,\mathbb{R})$ we have that ${\rm dim}\, G = n^2-1$ and ${\rm rank}\,
G=n-1$ so that we indeed verify the expression for $n_{\rm D}$ given
above. The other dimensions proceed in a similar way.

It is interesting to also consider the electric duals of the defect branes,
i.e.~instantons. These instantons occur in the same U-duality representations,
with the value of $\alpha$ given by the general relation
\begin{equation}
  \alpha_{\rm magnetic} = - \alpha_{\rm electric} -2 \quad,
\end{equation}
where in this case $\alpha_{\rm magnetic}$ is the value of $\alpha$ of a given
defect brane and $\alpha_{\rm electric}$ is the value of $\alpha$ of the dual
instanton. This implies that the values of $\alpha$ for instantons in $D\geq
3$ are $\alpha = 2,1,0,-1,-2$.  Since under S-duality the value of $\alpha$
transforms according to $\alpha\rightarrow\ -\,\alpha$ we see that the
instantons are symmetric around the $\alpha=0$ Fundamental instantons. The
Fundamental, Dirichlet and Solitonic instantons can all be understood as the
result of extending the corresponding wrapping rule to wrapping over time. For
instance, the Fundamental instantons arise as the result of applying the
fundamental wrapping rule (\ref{Fbranewrapping}) to the fundamental string,
see Table~\ref{7.5}. Note that fundamental instantons only arise in $3\le D\le
8$ dimensions.

\begin{table}[h]
\begin{center}
\begin{tabular}{|c||c|c|c|c|c|c|c|c|}
\hline \rule[-1mm]{0mm}{6mm} F$p$-brane &IIA/IIB& 9 & 8 & 7 & 6&5&4&3\\
\hline \hline \rule[-1mm]{0mm}{6mm} --1&&&4&12&24&40&60&84\\
\hline  \rule[-1mm]{0mm}{6mm} 0&&2&4&6&8&10&12&14\\
\hline \rule[-1mm]{0mm}{6mm} 1&1/1&1&1&1&1&1&1&1\\
\hline
\end{tabular}
\caption{\sl Upon applying the fundamental wrapping rule
(\ref{Fbranewrapping}) one obtains in each dimension the
U-duality representations of the Fundamental
instantons, cp.~to Table \ref{table2}.\label{7.5} }
\end{center}
\end{table}

The main result of our work is that we have associated a mixed-symmetry field
to each of the generalized KK monopoles using the general rule
(\ref{rule4}). All the generalized KK monopoles considered have a single set
of isometry directions, with the notable exception of the ten-dimensional
solution $(2,0,6,1)_{-4}$ and the eleven-dimensional solution $(2,0,7,1)$,
which have two {\sl inequivalent} isometry directions. Such monopoles have a
quadratic and cubic dependence of the mass on the radii, see eqs.~(\ref{mass})
and (\ref{mass2}).

It turns out that the specific mixed-symmetry fields we found are precisely
the ones predicted by ${\rm E}_{11}$ \cite{West:2001as}. Indeed, ${\rm
  E}_{11}$ naturally contains fields that are all possible dual descriptions
of the supergravity fields, and thus naturally includes the fields in Table
\ref{table3} and in eq.~(\ref{elevendimfieldsextable}) \cite{Riccioni:2006az}.
Moreover, selecting out of the various potentials the ones that are associated
to supersymmetric branes corresponds to selecting the real roots of ${\rm
  E}_{11}$, and this gives automatically all supersymmetric branes in all
dimensions \cite{Kleinschmidt:2011vu}.  This is one more application where
${\rm E}_{11}$ is used to learn about the properties of supergravity.

It is important to distinguish between the status of the mixed-symmetry fields
and that of the monopole solutions. The monopole solutions have been given in
the literature as solutions of the full non-linear supergravity theory
\cite{LozanoTellechea:2000mc,Englert:2007qb}.\,\footnote{It remains to be seen
  whether these monopole solutions can be turned into non-singular
  finite-energy solutions.} On the other hand the mixed-symmetry fields can
only be made consistent with supersymmetry at the level of {\sl linearized}
supersymmetry. A prime example is the dual graviton field $A_{8,1}$ in $D=11$
dimensions whose supersymmetry properties have been discussed in
\cite{Bergshoeff:2008vc}.  The restricted reduction rule of the mixed-symmetry
fields $A_{m,n}$ we found suggests that they couple to a generalized KK
monopole via a Wess-Zumino term where the last $n$ indices are taken into the
isometry directions and $n$ of the first $m$ indices are taken into the same
isometry directions. The remaining $m-n$ indices couple to the worldvolume
directions of the monopole in the usual way. It would be interesting to see
whether such a gauge-invariant WZ term describing the coupling of the
background fields to the monopole can be constructed.  \bigskip

\centerline{NOTE ADDED}
\bigskip

During the course of this work the paper \cite{Kleinschmidt:2011vu} appeared
which has some overlap with this work. In particular, Section 3 of
\cite{Kleinschmidt:2011vu} discusses defect brane solutions from an
$E_{11}$-point of view.


\section*{Acknowledgements}

E.B.~wishes to thank Axel Kleinschmidt for a useful correspondence.
T.O.~and~F.R.~wish to thank the Center for Theoretical Physics of the
University of Groningen for its hospitality and financial support.  The work
of T.O.~has been supported in part by the Spanish Ministry of Science and
Education grant FPA2009-07692, the Comunidad de Madrid grant HEPHACOS
S2009ESP-1473, and the Spanish Consolider-Ingenio 2010 program CPAN
CSD2007-00042. T.O.~wishes to thank M.M.~Fern\'andez for her unfaltering
support.


\vskip -20truecm






\end{document}